\documentclass[letterpaper, 10 pt, conference]{ieeeconf}  
\IEEEoverridecommandlockouts                                                           
\usepackage{cite}
\usepackage{amsmath}
\usepackage{graphicx}
\usepackage[font=footnotesize,labelfont=bf,skip=0pt]{caption}
\usepackage{epstopdf}
\usepackage{amssymb}
\usepackage{enumerate}
\usepackage{etaremune}
\usepackage{setspace}
\usepackage{color}
\usepackage{booktabs}
\usepackage{float}
\usepackage{pgfplots} 

\title{\LARGE \bf
An Audio-Based Fault Diagnosis Method for Quadrotors Using Convolutional Neural Network and Transfer Learning
}

\author{Wansong Liu$^{1}$, Zhu Chen$^{2}$, Minghui Zheng$^{3,*}$
	\thanks{$^{1,2,3}$ Wansong Liu, Zhu Chen, and Minghui Zheng are with the Mechanical and Aerospace Engineering Department, University at Buffalo, Buffalo, NY14260, USA.
		{\tt\small Emails: \{wansongl, zhuchen, mhzheng\}@buffalo.edu}.}%
		\thanks{$^*$ Corresponding Author.}
}

\begin{document}
\maketitle
\thispagestyle{empty}
\pagestyle{empty}
%%%%%%%%%%%%%%%%%%%%%%%%%%%%%%%%%%%%%%%%%%%%%%%%%%%%%%%%%%%%%%%%%%%%%%%%%%%%%%%%
\begin{abstract}
Quadrotor unmanned aerial vehicles (UAVs) have been developed and applied into several types of workplaces, such as warehouses, which usually involve human workers. The co-existence of human and UAVs brings new challenges to UAVs: potential failure of UAVs may cause risk and danger to surrounding human. Effective and efficient detection of such failure may provide early warning to the surrounding human workers and reduce such risk to human beings as much as possible. One of the \textcolor{black}{most common} reasons that cause the failure of the UAV's flight is the physical damage to the propellers. This paper presents a method to detect the propellers’ damage only based on the audio noise caused by the UAV's flight. The diagnostic model is developed based on convolutional neural network (CNN) and transfer learning techniques. The audio data is collected from the UAVs in real time, transformed into the time-frequency spectrogram, and used to train the CNN-based diagnostic model. The developed model is able to detect the abnormal features of the spectrogram and thus the physical damage of the propellers. To reduce the data dependence on the UAV's dynamic models and enable the utilization of the training data from UAVs with different dynamic models, the CNN-based diagnostic model is further augmented by transfer learning. As such, the refinement of the well-trained diagnostic model ground on other UAVs only requires a small amount of UAV's training data. Experimental tests are conducted to validate the diagnostic model with an accuracy of higher than 90\%.  
	\end{abstract}
%%%%%%%%%%%%%%%%%%%%%%%%%%%%%%%%%%%%%%%%%%%%%%%%%%%%%%%%%%%%%%%%%%%%%%%%%%%%%%%%

\section{Introduction}
\noindent
In recent years, considerable attention has been paid on unmanned aerial vehicles (UAVs) due to its broad application in many areas such as transportation monitoring \cite{barmpounakis2016unmanned}, infrastructure inspection \cite{sajedi2019convolutional,liang2019image}, and disaster resilience \cite{liang2018scalable,zheng2019preliminary}. Given the high risk of the pilot fatality, the early idea about UAV usage is reconnaissance missions in military around 1960 \cite{callam2010drone}. Semsch et al. \cite{semsch2009autonomous} presented an occlusion-aware mechanism to allow multiple UAVs to complete persistent surveillance in complex urban environments. Han et al. \cite{han2013low} utilized a bunch of low-cost UAVs to detect the nuclear radiation of a certain region periodically and efficiently. Abdelkader et al. \cite{abdelkader2013uav} combined a swarm of UAVs with Lagrangian microsensors and generated a real-time flood map with short-term flood propagation forecast to reduce casualties.

No matter in which field, the dynamic stability of UAVs is the premise of sophisticated scenarios. Therefore, the topic of the diagnosis on UAVs has attracted extensive research interest. Hansen et al. \cite{hansen2014diagnosis} applied parameter adaptive estimators to develop a diagnostic scheme to detect airspeed sensor fault. Avram et al. \cite{avram2015imu} developed a diagnostic scheme by estimating the roll and pitch angles ground on the inertial measurement units (IMUs) of UAVs. Younes et al. \cite{al2016sensor} introduced an output-estimator  to detect the sensor fault. The signals from the fault sensor could affect some computational commands of the actuators.  

Meanwhile, compared to the UAV sensor failure, the physical damage to the propeller could bring more serious consequences. UAVs might lose the flying balance when the damaged propeller rotation cannot maintain the desired lift force, and fall directly. The crucial concern that is human beings' life safety is threatened by this physical failure. Benini et al. \cite{benini2019fault} developed a damaged propeller blade detection model based on the acceleration signals from the IMU of a UAV. Ghalamchi et al. \cite{ghalamchi2018vibration} identified the broken propeller by detecting the physical damage vibration data provided by a built-in accelerometer. Rangel-Magdaleno et al. \cite{de2018detection} implemented Discrete Wavelet Transform (DWT) decomposition and Fourier Transform to process audio data collected from a real experimental test, and used some statistical parameters based on audio data to discriminate the broken propeller. 

The audio data indicates the noise amplitude of UAVs, and there are several approaches to analyze the abnormal behavior of UAVs based on it. One \textcolor{black}{common} one is spectrogram. Harmanny et al. \cite{harmanny2015radar} utilized spectrogram to distinct birds and mini-UAVs based on the radar micro-Doppler. While even it is possible to detect the broken propellers from the spectrogram of the vibrations by human, the efforts on some tremendous amount of works still could be a huge burden of human beings. In terms of accuracy and efficiency, deep learning techniques have started to show great potential in the fault detection areas \cite{iannace2019fault,zhang2018deep,chen2017deep,de2016railway}. The CNN has shown great success in making decisions on image-based studies \cite{pereira2016brain,levi2015age,ma2016learning}. Researchers define the CNN code to split one image into several parts and extract some particular features. Hence, CNN can identify the distinction remarkably on image-based fault diagnosis. Wen et al. \cite{wen2017new} proposed an approach to transfer the features of the raw data from datasets to two-dimensional images, and utilized a classical release of CNN named LeNet-5 to do fault diagnosis. Guo et al. \cite{guo2018hybrid} converted the residual signals from global positioning system (GPS), IMU and air data system (ADS) to time-frequency maps, and implemented the CNN to diagnose damaged UAVs ground on the extracted features of 2D maps. 

\begin{figure}
	\centering 
	\includegraphics[scale=0.45]{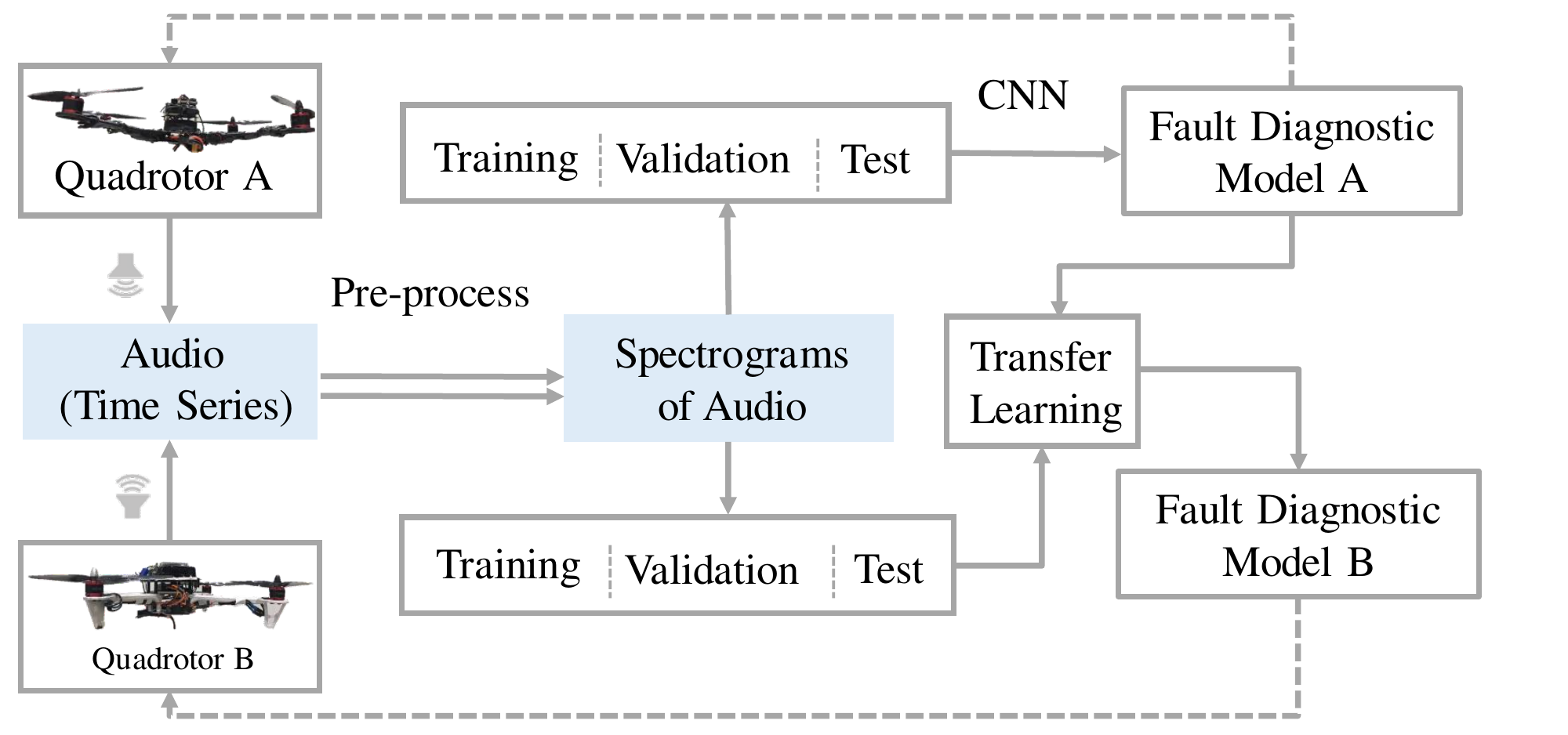}
	\caption{Framework of proposed diagnosis method } \label{fig:paper_structure}
\end{figure}

Considering the trained neural network usually only shows decent performance on one specific model, it is desirable to exploit the generalization of fault diagnosis according to a neural network. In the past decade, transfer learning has been spotlighted as a novel learning technique for extending the adaption of a model. Wen et al. \cite{wen} presented a supervised transfer learning based on sparse auto-encoder, and diagnosed fault by minimizing the discrepancy penalty between two tasks.  Zhang et al. \cite{zhang2017transfer} utilized transfer learning to improve bearing fault diagnosis performance in different working environments. Cao et al. \cite{cao2018preprocessing} realized gear transmission diagnosis with a small set of training data by transferring the knowledge from a well-trained neural network. 

This study aims to provide a new yet easy deep learning method for the UAVs' propeller fault diagnosis based on the 2D audio spectrogram. The audio data is collected from several actual UAVs. The traditional audio-based fault diagnosis which directly uses audio signal data to do analysis \cite{glowacz2018early,kemalkar2016engine}, while this paper converts the collected audio data to a time-frequency map that informs the audio energy at a certain time with a specific frequency by the spectrogram. CNN classifies the broken and unbroken propellers
ground on using spectrograms as the input data. Additionally, transfer learning is employed to build up a generalized model for different UAVs with various testing scenarios. The results show this approach has an outstanding capability for the UAVs' propeller fault diagnosis. 

The remainder of this paper is organized as follows. Section II introduces the whole framework; Section III presents the data collection and pre-process; Section IV presents the diagnosis method based on CNN; Section V extends this method to different UAVs based on transfer learning; Section VI concludes this paper.

\section{Proposed Diagnosis Scheme}
\noindent
Physical damage to the propellers of the quadrotor may cause unsteady flight or even flight failure. While several diagnosis techniques have been developed based on the control signals or the accelerations of the quadrotors as well as the dynamics, in this paper we present a model-free diagnosis method that is only based on the noise introduced by flight. It is easy to be implemented and does not require any additional sensors. 
	
The overall scheme of the proposed diagnosis method, as illustrated in Fig.~\ref{fig:paper_structure}, includes the following key components: audio recorder to obtain the time-domain audio signals, spectrogram converter to transform the signals from the time domain to the time-frequency domain, CNN-based diagnostics, and transfer learning-based diagnostics. The audio data from Quadrotor A is received by the audio recorder in terms of time series, and then transferred to the ``spectrogram'' in the time-frequency domain which can be presented by two-dimensional images. These images are labeled by the scenarios indicating whether damaged propellers are included or not and used to train or test the diagnostics model using CNN. To make the trained CNN model to be extensible to other quadrotors, transfer learning is used to refine the CNN model when the audio source becomes a different quadrotor, denoted as Quadrotor B.

\section{Audio Data Collection and Analysis}
\subsection{Audio data processing}
The audio data indicates the amplitude of the noise along the time interval from UAVs. While there are still some audio characteristics that show in the frequency domain. Thus, this study uses a comprehensive presentation, ``spectrogram'', which represents the audio signal features in both time and frequency domains. The audio data is transferred to a series of windowed segments, and the windowed segments are offset adjacently in the time domain with a certain percent.  In a spectrogram, the time and frequency are represented along the horizontal axis and vertical axis, respectively. Therefore, the audio signal is shown in the spectrogram continuously, and the energy of the audio signal can be denoted as different colors at a particular time with a certain frequency.  At each time instance $t$ and frequency $k$, the spectrogram $X(t, k)$ in the discrete-time domain is denoted as: 
\begin{equation}
x(t, k)=\sum_{n=0}^{N-1} W(n) \cdot x(n+t\cdot h) e^{-2 i \pi \frac{k}{N} n}
\end{equation}
where $w(n)$ is a Hann window with $N$ samples, $x(t)$ is the audio signal, and $h$ is the time shift between the continuous windowed segments.

\vspace{-10pt}
\begin{figure}[!htbp]
	\centering 
	\includegraphics[scale=0.45]{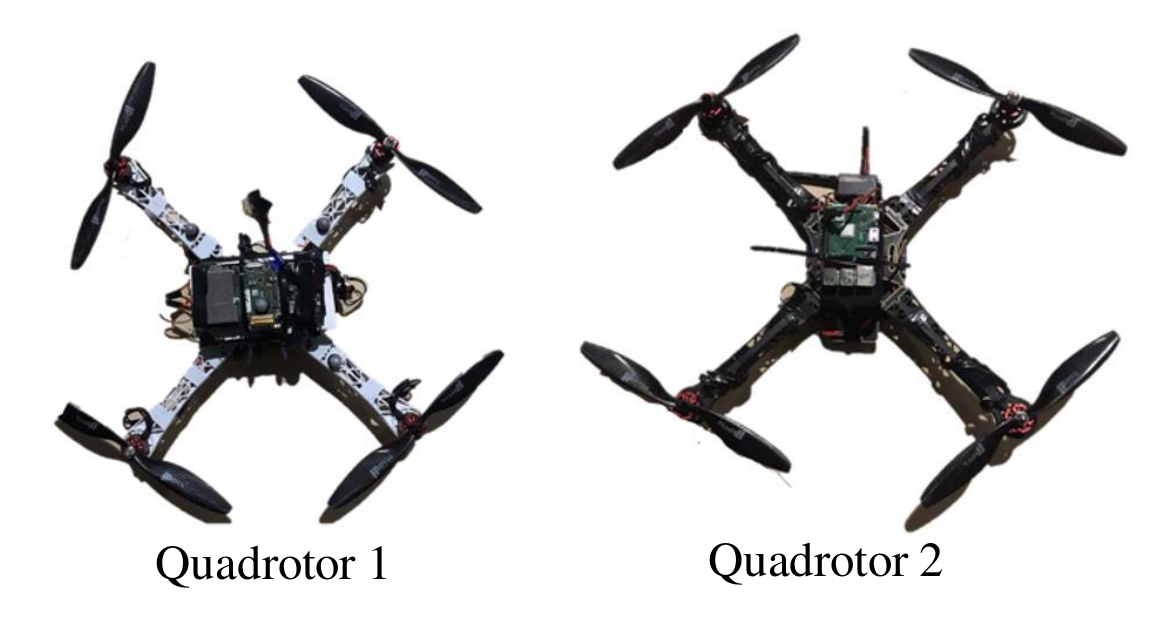}
	\vspace{0pt}
	\caption{Quadrotors used for experimental test} \label{fig:uav1}
\end{figure}

\vspace{-20pt}
\begin{figure}[!htbp]
	\centering 
	\includegraphics[scale=0.45]{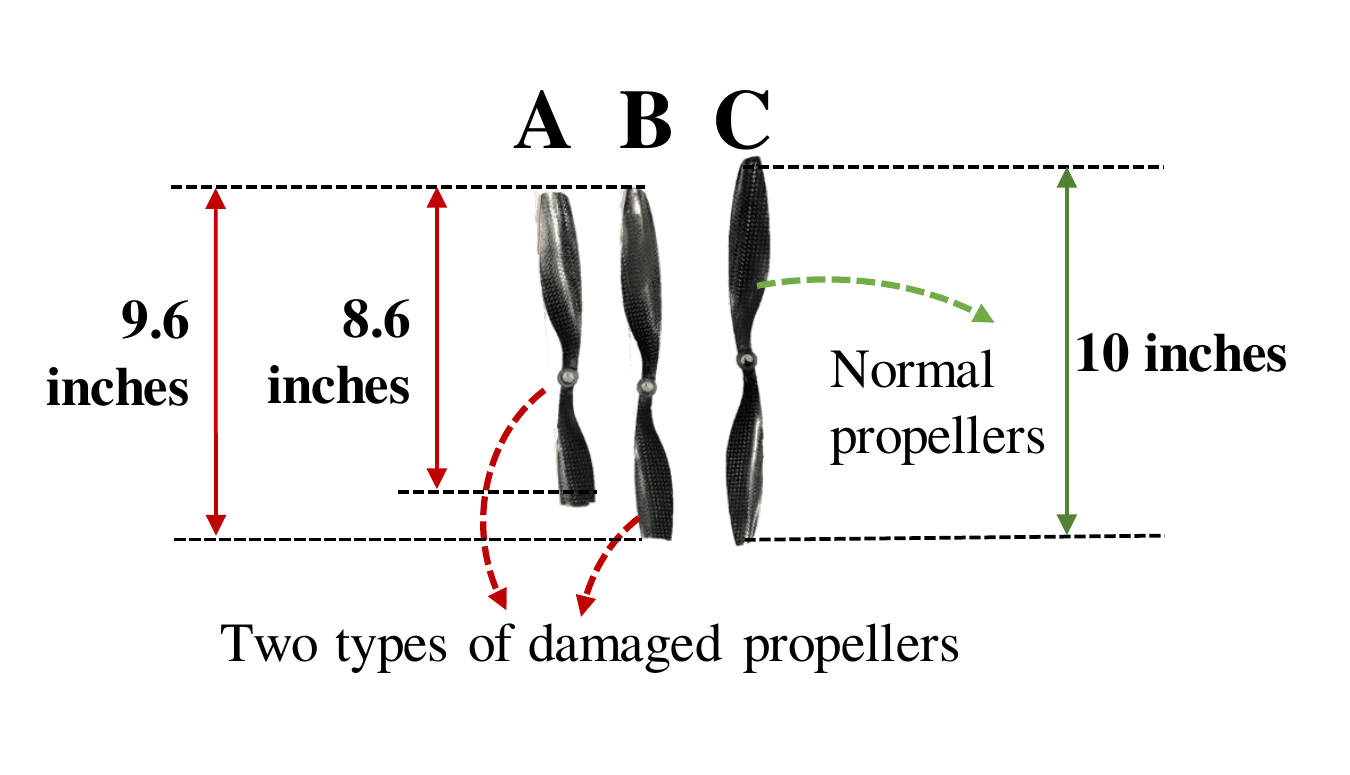}
		\vspace{-5pt}
	\caption{Propellers used for experimental test} \label{fig:uav2}
\end{figure}

\begin{table*}[t]
	\centering
	\begin{tabular}{l|c|c|c|c|c}
\toprule[1.5pt]		Tests & Propeller Sets & Thrust Commands \\\toprule[1.5pt]
		Configuration 1 & Four propellers in good condition & Low/medium/high\\
		\hline
		Configuration 2 & Three propellers in good condition plus propeller A & Low/medium/high\\
		\hline
		Configuration 3 & Three propellers in good condition plus propeller B & Low/medium/high \\
		\toprule[1.5pt]
	\end{tabular}
	\vspace{5pt}
	\caption{Test configurations}
	\label{tab:test}
\end{table*}

\subsection{Experimental test platform and data collection}
The platform mainly consists of quadrotors, an audio recorder, a computer, and several types of propellers with different levels of damages. In this paper, two assembled quadrotors with different frames, as shown in Fig.~\ref{fig:uav1}, are used for the data collection and experimental test. Well-balanced propellers and damaged propellers are shown in Fig.~\ref{fig:uav2}. The propeller labeled with “C” in Fig.~\ref{fig:uav2} is in good condition and has a dimension of 10 inches for diameter and 4.5 inches for pitch. The length of the propeller labeled with “A” is 8.6 inches and the length of the propeller labeled with “B” is 9.6 inches, which illustrates the defections for propellers A and B.

\begin{figure}[!htbp]
	\centering 
	\includegraphics[scale=0.4]{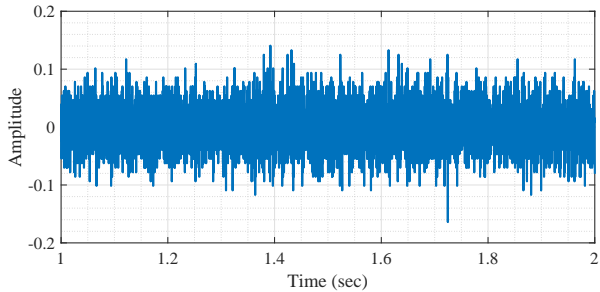}
	\vspace{-1pt}
	\caption{Audio signals generated from the test with broken propeller} \label{fig:audio_broken}
\end{figure}

\begin{figure}[!htbp]
	\centering 
	\includegraphics[scale=0.4]{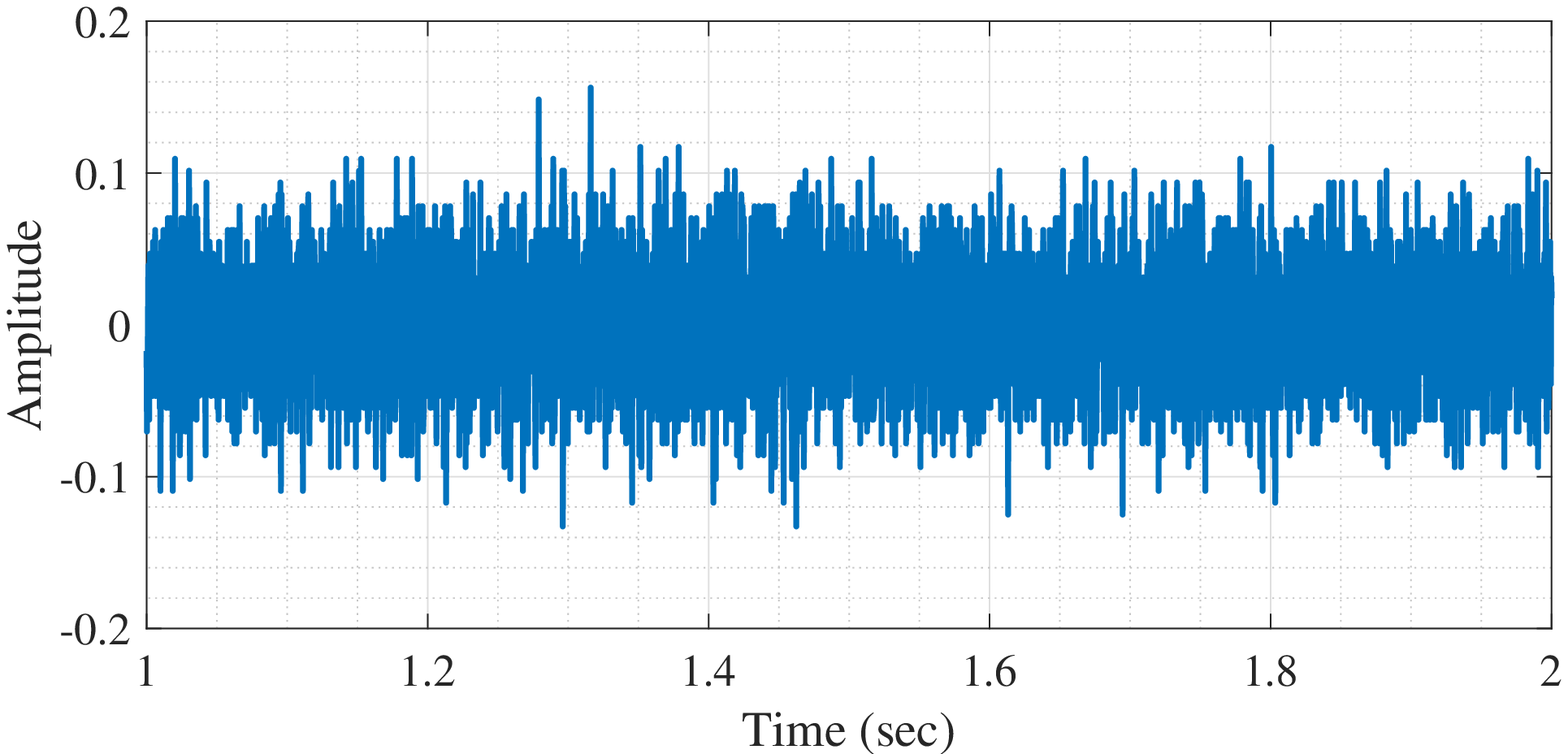}
	\vspace{2pt}
	\caption{Audio signals generated from the test with unbroken propeller}
		\vspace{5pt}
	\label{fig:audio_unbroken}
\end{figure}

\begin{figure}[!htbp]
	\centering 
	\includegraphics[scale=0.45]{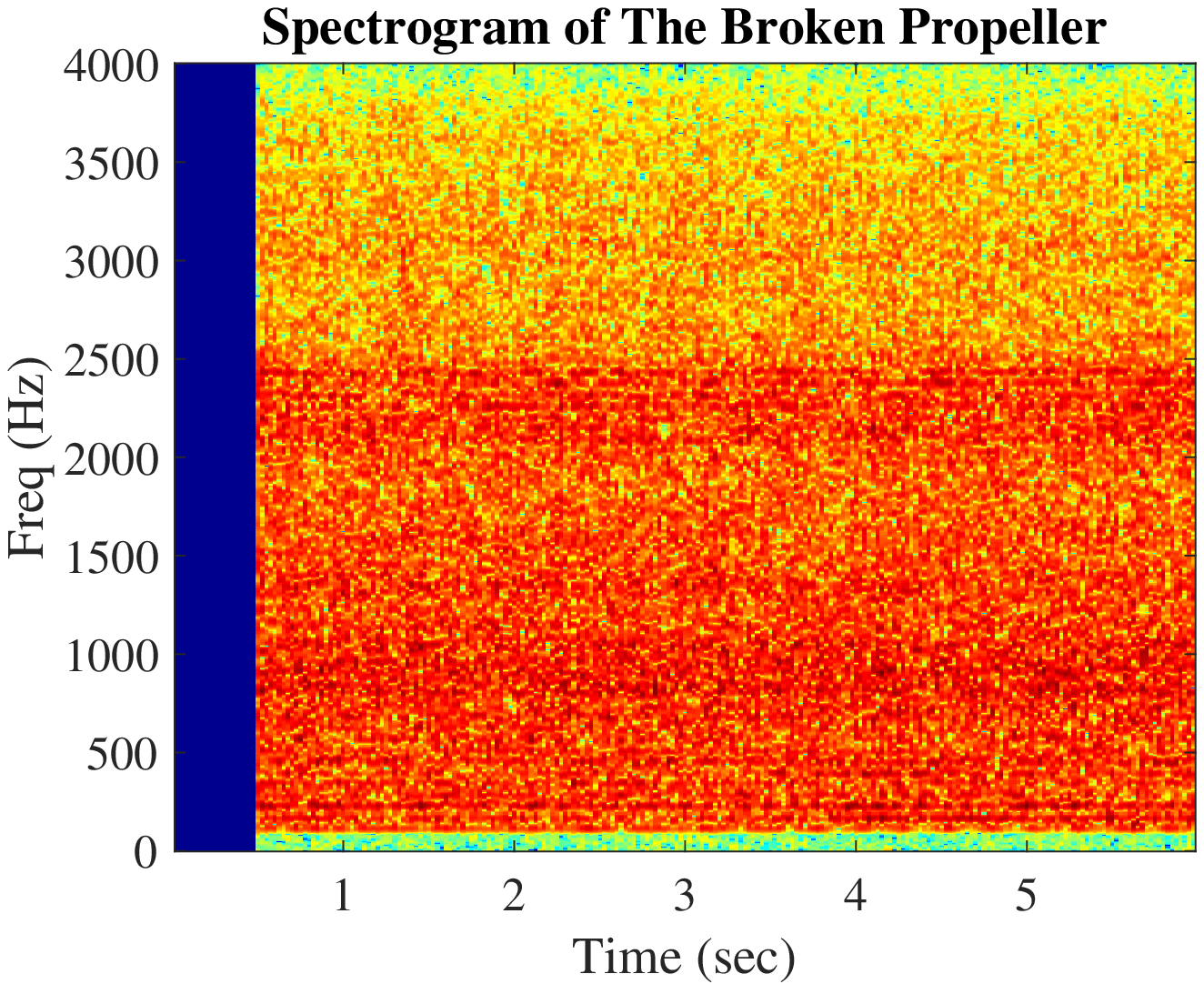}
	\vspace{10pt}
	\caption{Spectrogram generated from the test with broken propeller} \label{fig:s_broken1}
\end{figure}

\begin{figure}[!htbp]
	\centering 
	\includegraphics[scale=0.45]{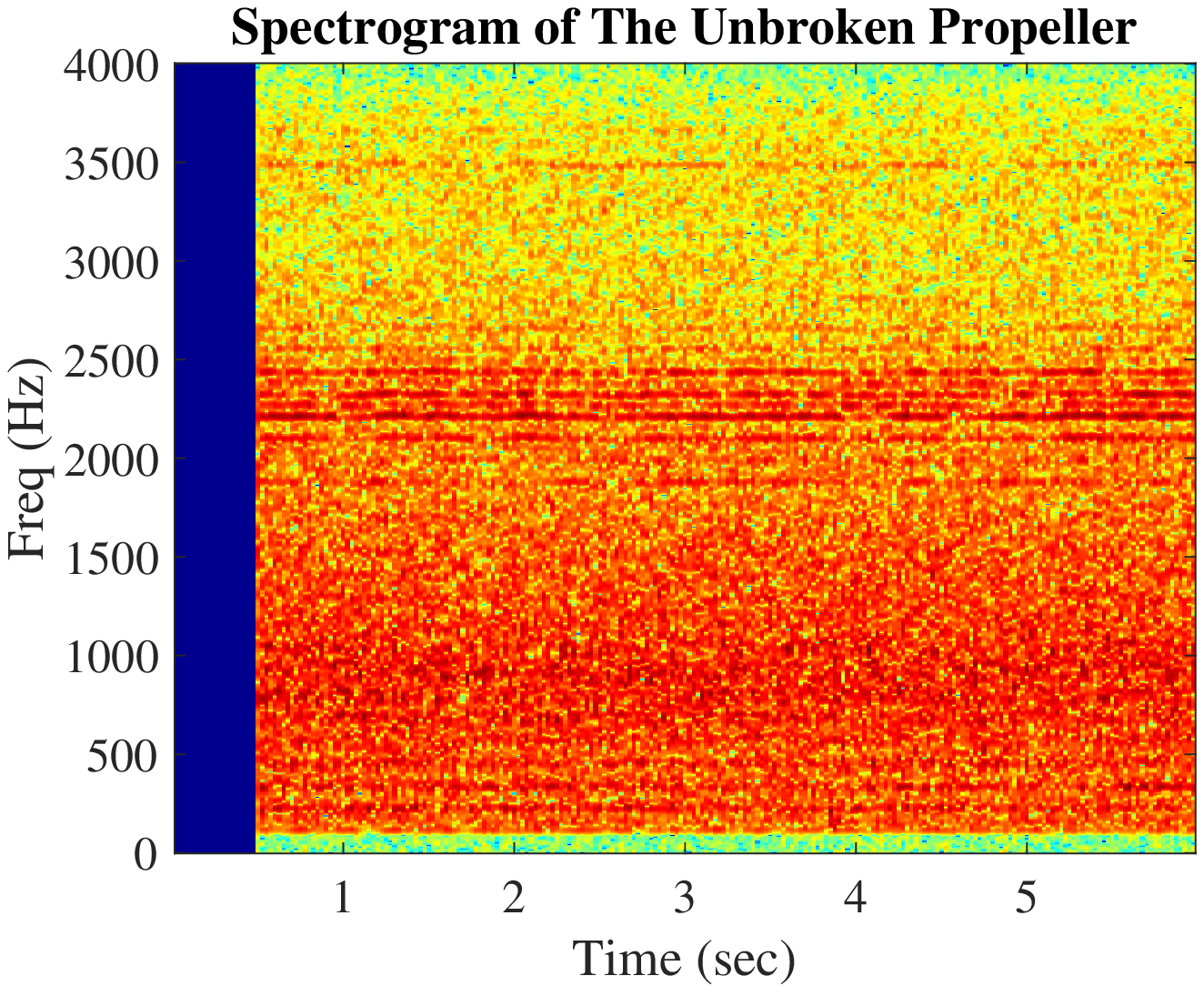}
	\vspace{10pt}
	\caption{Spectrogram generated from the test with unbroken propeller}
	\vspace{-10pt}\label{fig:s_unbroken1}
\end{figure}

\begin{figure*}[t]
	\centering 
	\includegraphics[scale=0.5]{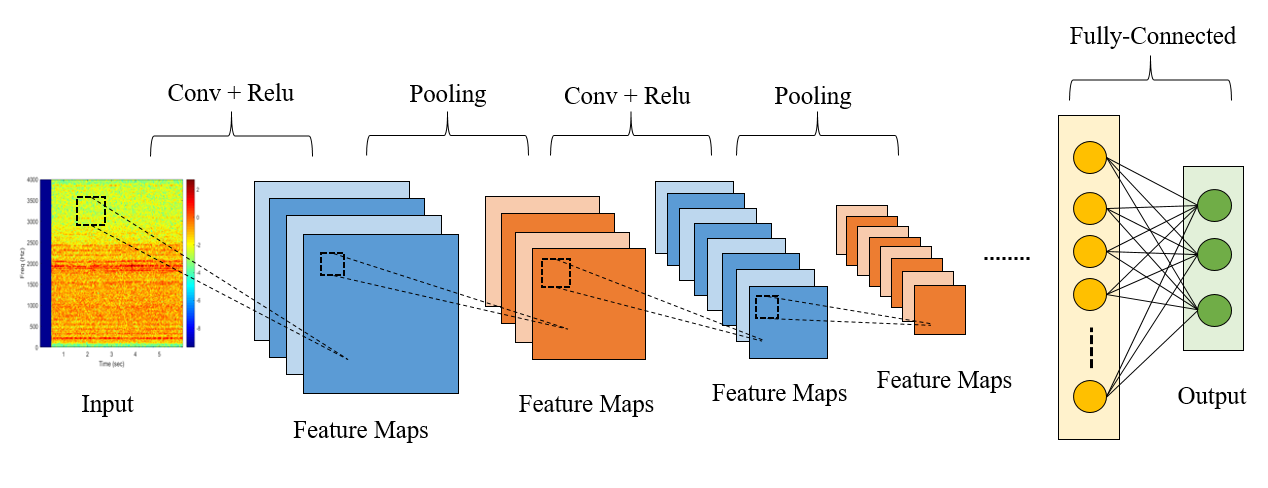}
	\vspace{-10pt}
	\caption{The structure of CNN} \label{fig:CNN}
\end{figure*}

Three sets of tests are conducted for each quadrotor. Each set of test is with different propeller configuration. In order to obtain the quadrotors' noise data in different speed scenarios, different thrust commands are applied for each set of tests as shown in Table.~\ref{tab:test}. Three sets of thrust commands which are defined as low, medium, and high thrust commands, are given through the radio control with a frequency of 2.4 GHz. A microphone records the audio with a duration of 6 seconds as a sample of data. Data samples are presented as the form in Fig.~\ref{fig:audio_broken} and Fig.~\ref{fig:audio_unbroken}.

Considering the propellers of UAVs are in a high rotation velocity, the audio signal generated from the vibration may display some particular features in the spectrogram based on the broken propeller's unbalance. Fig.~\ref{fig:s_broken1} and Fig.~\ref{fig:s_unbroken1} show the spectrograms of the audio data collected from the broken and unbroken propellers in the same scenario, separately. The audio signal distributes successively along the frequency and time axis with diverse colors due to the energy variation. The audio data emerges as a straight line with several certain frequency channels in the horizontal direction. Thus, the integrality and consecutiveness of the audio data from UAVs have been demonstrated in the spectrogram outstandingly.

\section{Image-Based Deep Learning Technique}
\subsection{Introduction to CNN}
Despite the spectrogram presents the audio data characterization promisingly, human beings still need an automatic way to distinguish the spectrogram-based features of the broken propeller when facing a tremendous workload. Deep learning techniques including CNNs have been used for automated health monitoring of different systems including infrastructure \cite{sajedi2019intensity,sajedi2019vibration,liang2018simulation,sajedi2020data}. This paper utilizes CNN to detect the difference between broken and unbroken propellers. The spectrograms of the audio data collected from UAVs are employed as the CNN training object. The spectrograms are segregated to the training data, validation data and test data. The training data is the input of the neural network, and it takes the most proportion of the whole dataset. The validation data does not go through the neural network, and it is applied to verify the classification quality of the neural network model and avoid overfitting during the training process. The test data is utilized to evaluate the performance of the well-trained neural network. Fig.~\ref{fig:CNN} shows the \textcolor{black}{general} architecture of CNN.

The spectrogram as input data has three digital channels. Each channel is represented by a particular matrix with the pixel value range from 0 to 255. Another matrix called kernel (filter) slides over each channel with a certain stride value, and gets the dot product to construct feature maps. The Relu is a kind of activation function to set all negative value to zero. Considering the stride value is usually small, the dimension of the feature maps will stretch over and over. In other words, the tremendous computation of the training process turns to be a burden of the classification efficiency.  Therefore, CNN employs a method called pooling to diminish the dimension of feature maps. Pooling separates the feature map to several spatial continuous matrices with the same dimension, and selects the maximum element of each matrix or the average value of elements from each matrix. After the feature extraction of the input data, the fully connected part transfers the feature compositions to a more straightforward way. This paper applies Softmax function to turn extracted features to a series of possible scores from 0 to 1. 
At the beginning of the training progress, CNN will set the initial parameter arbitrarily. After a forward propagative iteration, the output is denoted as the following equation:
\begin{equation}
Y=f(w\cdot x+b)
\end{equation}
where $x$ is a series of input, $f$ is an activation function, $w$ and $b$ are the weight and bias terms during the training. Meanwhile, the neural network calculates the output error backpropagatively, and updates the weights in each layer. 

\begin{figure}[!htbp]
	\centering 
	\includegraphics[scale=0.45]{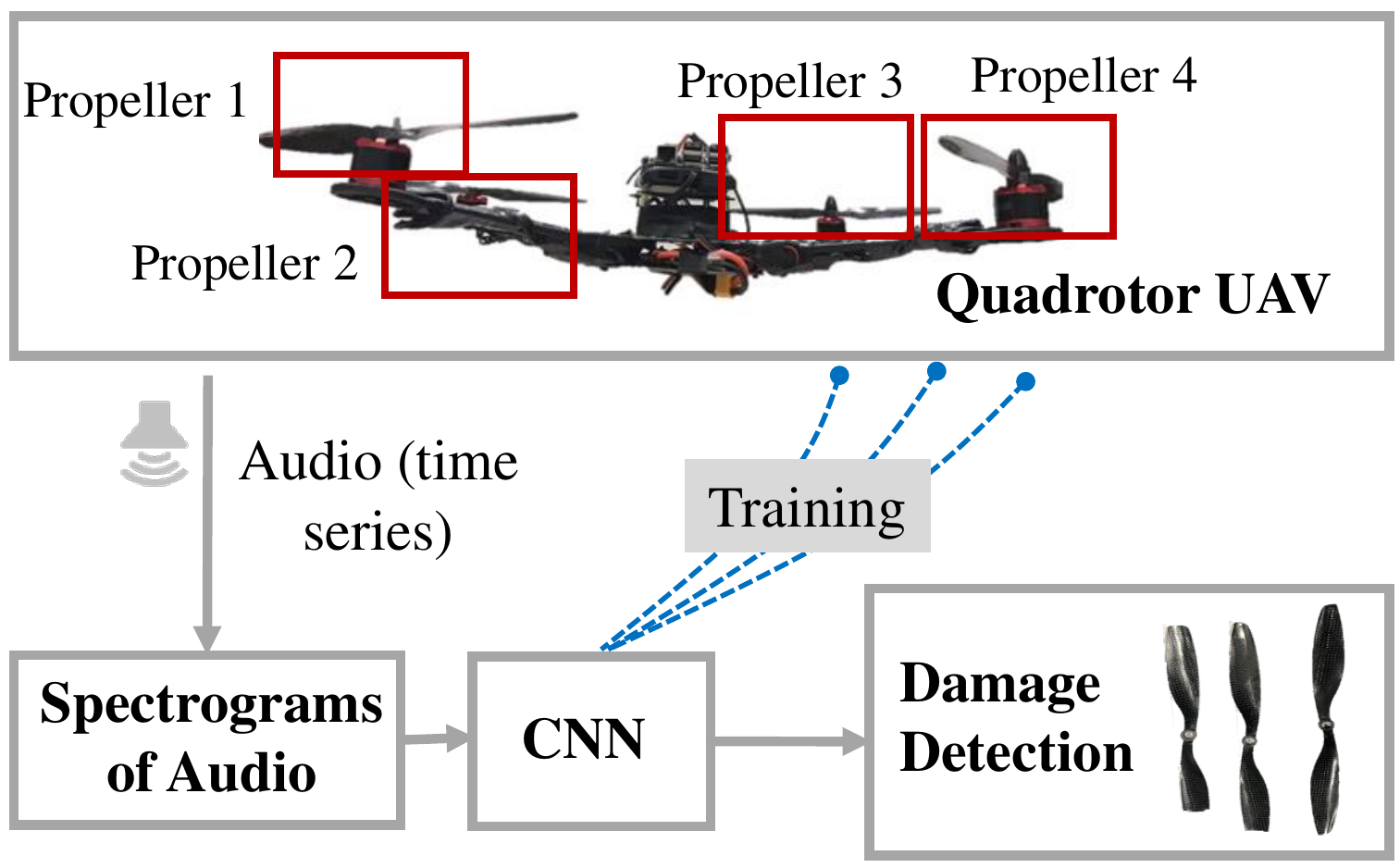}
		\vspace{10pt}
	\caption{The structure of the experimental test to train and validate CNN} \label{fig:CNN_structure}
\end{figure}

\subsection{Image classification test}

Fig.~\ref{fig:CNN_structure} presents the framework of the experimental test based on the spectrogram and CNN. The audio signals generated by four propellers' rotation in different scenarios are converted to 2D images. The CNN \textcolor{black}{which contains one convolutional layer }treats spectrograms of audio as input data to detect the damaged propellers.   

\begin{figure}[!htbp]
	\centering 
	\includegraphics[scale=0.3]{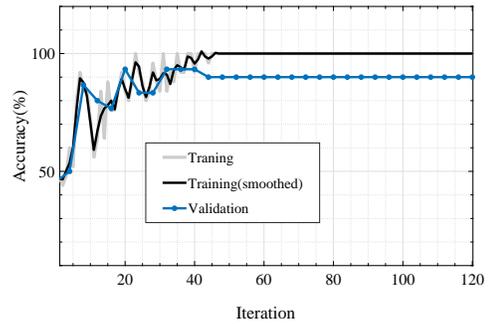}
	\vspace{1pt}
	\caption{Training and validating of the CNN} \label{fig:network}
\end{figure}
Fig.~\ref{fig:network} shows the training progress of CNN. The total 160 spectrogram images are comprised of 80 images from the broken propeller and 80 images from the unbroken propeller, and lay in two different files, respectively. The Matlab code randomly selects 50 images from each file to compose the training data (100 images), and randomly selects 15 images from each file to compose the validation data (30 images). The remaining images make up the test data (30 images). Since there are only two kinds of outputs, the classification accuracy begins around 50\%. After the network training, the classification from the neural network is extremely efficient and accurate, and the test data accuracy can be 96.67\%.  All the spectrograms are collected in the different propeller rotation speed, thus this approach could avoid the damage caused by the propeller unbalance, especially when the rotation speed is relatively low and UAVs have not taken off from the ground. On the other hand, if UAVs are flying aerially, the fault detection can allow UAVs to take measurements quickly and escape some further damages.

\section{Generalized Diagnostic Model for UAVs}
\subsection{CNN-based transfer learning technique}
The CNN with the spectrogram has a striking performance on the UAV propeller fault diagnosis. Given different UAVs have various characteristics, like the propeller size, the frame of the UAV, the propeller rotation speed and so on, it's not universal to apply a well-trained neural network on all kinds of UAVs. Therefore, each UAV has to develop its own neural network to detect failures. However, numerous training data requirement is one of the essential properties of the neural network, which constrains the generalization of this fault diagnosis approach. Thus, this paper implements a method called transfer learning to transfer the knowledge from a well-trained UAV model to a new UAV model with insufficient data. 
The definition of transfer learning refers to the work from Pan and Yang (2010) \cite{pan2009survey}. Pan and Yang introduced two conceptions called “domain” and “task”. For the domain, it is denoted as $D$ and comprised of a feature space $X$ and a marginal probability distribution $P(x)$. For the task, it is denoted as $T$ and comprised of a label space $y$ and an objective predictive function $f(\cdot)$ that is written as $P(y/x)$. Transfer learning utilizes the knowledge ground on a source domain $D_S$ and learning task $T_S$ to improve the target predictive function $f(\cdot)$ performance of a learning task $T_T$ in a target domain $D_T$, where $D_S \neq D_T$, $T_S \neq T_T$. 

\subsection{Transfer learning experimental test and improvement}
\begin{figure}[ht]
	\centering 
	\includegraphics[scale=0.45]{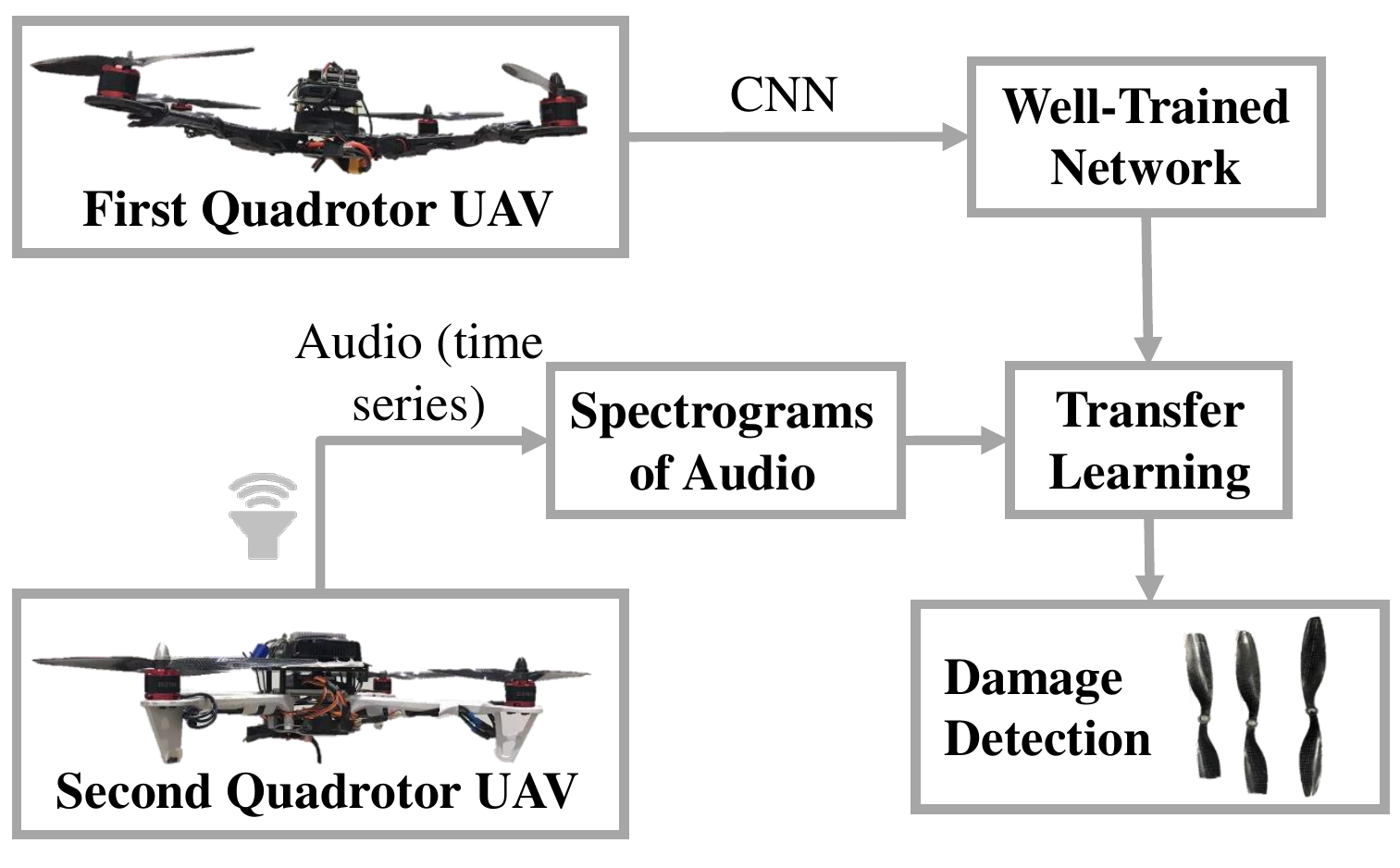}
		\vspace{10pt}
	\caption{Transfer learning framework} \label{fig:TF_structure}
\end{figure}

\begin{figure}[ht]
\vspace{-10pt}
	\centering 
	\includegraphics[scale=0.35]{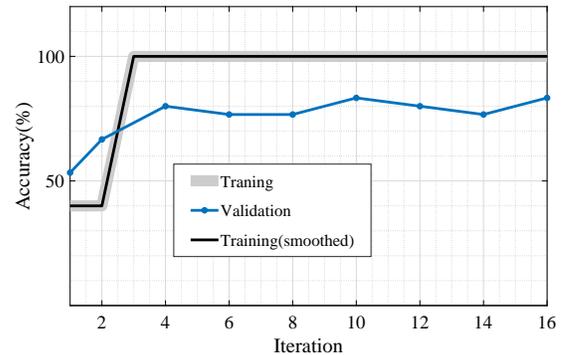}\vspace{-5pt}
	\caption{Training progress based on transfer learning with 10 training images} \label{fig:transfer_learning_10}
\end{figure}

\begin{figure}[ht]
	\centering 
	\includegraphics[scale=0.35]{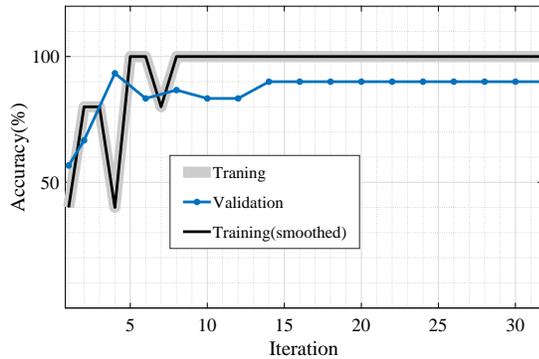}\vspace{-5pt}
	\caption{Training progress based on transfer learning with 20 training images} \label{fig:transfer_learning_20}
\end{figure}

Fig.~\ref{fig:TF_structure} shows the framework of transfer learning experimental test. 
The abundant audio training data from the first quadrotor is employed by CNN to set up a well-trained network. The Matlab code utilizes most network layers and isolates the last few layers of the network, and replaces them by new layers that learn specific features based on the spectrograms of audio from the second quadrotor. The transfer learning only needs a few training data from the second quadrotor, and fine-tunes the weights of each layer during the training progress. 
The spectrogram images with the same total number are collected from the second quadrotor. As this paper mentioned before, the classification accuracy of the first quadrotor is 96.67\%. We apply all the spectrogram images collected from the second quadrotor as test data on the well-trained diagnostic network model. The classification accuracy without transfer learning is 55.00\% which almost corresponds to the proportion of the propeller types. To improve the generalization of the fault diagnostic model, the Matlab code randomly selects 5 spectrogram images from the broken and unbroken propeller files, respectively. The 10 spectrogram images compose the training data of the transfer learning progress. The validation data is still made up by 30 spectrogram images. The remaining 120 spectrogram images are the test data. Fig.~\ref{fig:transfer_learning_10} indicates the training progress, and the classification accuracy of the test data from the second quadrotor is 81.67\% which is much higher than 55.00\%. To increase the classification accuracy, we modify the number of training spectrogram images to be 20. The validation data is still made up by 30 spectrogram images. The remaining 110 spectrogram images are the test data. Fig.~\ref{fig:transfer_learning_20} indicates the training progress of transfer learning with 20 training images, and the classification accuracy is 91.82\%. Hence, as different quadrotors, even though the parameters of quadrotors are various, the audio-based fault diagnosis model  still shows a promising accuracy ($91.82\%$).

\vspace{-20pt}
%\clearpage
\vspace{-0pt}
\section{Conclusions}
This paper presents a new fault diagnostic method for physical damage of propellers used on quadrotor UAVs. The method only needs the audio data from the quadrotor's flight and does not require the physical model of the quadrotor. This diagnostic model is first developed based on the CNN, which is trained by labeled data gathered from quadrotors with different damaged propellers. The model is then generalized by transfer learning, such that the model trained by the data from one quadrotor can be applied to another quadrotor without too much additional training. One advantage of this method is that it is only based on audio data which is easy to obtain. The audio data is converted to the 2D spectrogram and sent to the CNN in terms of images. Experimental studies have been conducted to validate the effectiveness of the proposed fault diagnostic model.

\bibliographystyle{IEEEtran}
\bibliography{ref}{}

% Generated by IEEEtran.bst, version: 1.13 (2008/09/30)
\begin{thebibliography}{10}
\providecommand{\url}[1]{#1}
\csname url@samestyle\endcsname
\providecommand{\newblock}{\relax}
\providecommand{\bibinfo}[2]{#2}
\providecommand{\BIBentrySTDinterwordspacing}{\spaceskip=0pt\relax}
\providecommand{\BIBentryALTinterwordstretchfactor}{4}
\providecommand{\BIBentryALTinterwordspacing}{\spaceskip=\fontdimen2\font plus
\BIBentryALTinterwordstretchfactor\fontdimen3\font minus
  \fontdimen4\font\relax}
\providecommand{\BIBforeignlanguage}[2]{{%
\expandafter\ifx\csname l@#1\endcsname\relax
\typeout{** WARNING: IEEEtran.bst: No hyphenation pattern has been}%
\typeout{** loaded for the language `#1'. Using the pattern for}%
\typeout{** the default language instead.}%
\else
\language=\csname l@#1\endcsname
\fi
#2}}
\providecommand{\BIBdecl}{\relax}
\BIBdecl

\bibitem{barmpounakis2016unmanned}
E.~N. Barmpounakis, E.~I. Vlahogianni, and J.~C. Golias, ``Unmanned aerial
  aircraft systems for transportation engineering: Current practice and future
  challenges,'' \emph{International Journal of Transportation Science and
  Technology}, vol.~5, no.~3, pp. 111--122, 2016.

\bibitem{sajedi2019convolutional}
{S. O. Sajedi and X. Liang}, ``A convolutional cost-sensitive crack
  localization algorithm for automated and reliable {RC} bridge inspection,''
  in \emph{Risk-Based Bridge Engineering: Proceedings of the 10th New York City
  Bridge Conference, August 26-27, 2019, New York City, USA}.\hskip 1em plus
  0.5em minus 0.4em\relax CRC Press, 2019, p. 229.

\bibitem{liang2019image}
X.~Liang, ``Image-based post-disaster inspection of reinforced concrete bridge
  systems using deep learning with bayesian optimization,''
  \emph{Computer-Aided Civil and Infrastructure Engineering}, vol.~34, no.~5,
  pp. 415--430, 2019.

\bibitem{liang2018scalable}
X.~Liang, M.~Zheng, and F.~Zhang, ``A scalable model-based learning algorithm
  with application to uavs,'' \emph{IEEE Control Systems Letters}, 2018.

\bibitem{zheng2019preliminary}
M.~Zheng, Z.~Chen, and X.~Liang, ``A preliminary study on a physical model
  oriented learning algorithm with application to uavs,'' in \emph{ASME 2019
  Dynamic Systems and Control Conference}.\hskip 1em plus 0.5em minus
  0.4em\relax American Society of Mechanical Engineers Digital Collection,
  2019.

\bibitem{callam2010drone}
A.~Callam, ``Drone wars: Armed unmanned aerial vehicles,'' \emph{International
  Affairs Review}, vol.~18, no.~3, 2010.

\bibitem{semsch2009autonomous}
E.~Semsch, M.~Jakob, D.~Pavlicek, and M.~Pechoucek, ``Autonomous uav
  surveillance in complex urban environments,'' in \emph{Proceedings of the
  2009 IEEE/WIC/ACM International Joint Conference on Web Intelligence and
  Intelligent Agent Technology-Volume 02}.\hskip 1em plus 0.5em minus
  0.4em\relax IEEE Computer Society, 2009, pp. 82--85.

\bibitem{han2013low}
J.~Han, Y.~Xu, L.~Di, and Y.~Chen, ``Low-cost multi-uav technologies for
  contour mapping of nuclear radiation field,'' \emph{Journal of Intelligent \&
  Robotic Systems}, vol.~70, no. 1-4, pp. 401--410, 2013.

\bibitem{abdelkader2013uav}
M.~Abdelkader, M.~Shaqura, C.~G. Claudel, and W.~Gueaieb, ``A uav based system
  for real time flash flood monitoring in desert environments using lagrangian
  microsensors,'' in \emph{2013 International Conference on Unmanned Aircraft
  Systems (ICUAS)}.\hskip 1em plus 0.5em minus 0.4em\relax IEEE, 2013, pp.
  25--34.

\bibitem{hansen2014diagnosis}
S.~Hansen and M.~Blanke, ``Diagnosis of airspeed measurement faults for
  unmanned aerial vehicles,'' \emph{IEEE Transactions on Aerospace and
  Electronic Systems}, vol.~50, no.~1, pp. 224--239, 2014.

\bibitem{avram2015imu}
R.~C. Avram, X.~Zhang, J.~Campbell, and J.~Muse, ``Imu sensor fault diagnosis
  and estimation for quadrotor uavs,'' \emph{IFAC-PapersOnLine}, vol.~48,
  no.~21, pp. 380--385, 2015.

\bibitem{al2016sensor}
Y.~Al~Younes, A.~Rabhi, H.~Noura, and A.~El~Hajjaji, ``Sensor fault diagnosis
  and fault tolerant control using intelligent-output-estimator applied on
  quadrotor uav,'' in \emph{2016 International Conference on Unmanned Aircraft
  Systems (ICUAS)}.\hskip 1em plus 0.5em minus 0.4em\relax IEEE, 2016, pp.
  1117--1123.

\bibitem{benini2019fault}
A.~Benini, F.~Ferracuti, A.~Monteri{\`u}, and S.~Radensleben, ``Fault detection
  of a vtol uav using acceleration measurements,'' in \emph{2019 18th European
  Control Conference (ECC)}.\hskip 1em plus 0.5em minus 0.4em\relax IEEE, 2019,
  pp. 3990--3995.

\bibitem{ghalamchi2018vibration}
B.~Ghalamchi and M.~Mueller, ``Vibration-based propeller fault diagnosis for
  multicopters,'' in \emph{2018 International Conference on Unmanned Aircraft
  Systems (ICUAS)}.\hskip 1em plus 0.5em minus 0.4em\relax IEEE, 2018, pp.
  1041--1047.

\bibitem{de2018detection}
J.~de~Jesus Rangel-Magdaleno, J.~Ure{\~n}a-Ure{\~n}a, A.~Hern{\'a}ndez, and
  C.~Perez-Rubio, ``Detection of unbalanced blade on uav by means of audio
  signal,'' in \emph{2018 IEEE International Autumn Meeting on Power,
  Electronics and Computing (ROPEC)}.\hskip 1em plus 0.5em minus 0.4em\relax
  IEEE, 2018, pp. 1--5.

\bibitem{harmanny2015radar}
R.~I. Harmanny, J.~J. de~Wit, and G.~Premel-Cabic, ``Radar micro-doppler
  mini-uav classification using spectrograms and cepstrograms,''
  \emph{International Journal of Microwave and Wireless Technologies}, vol.~7,
  no. 3-4, pp. 469--477, 2015.

\bibitem{iannace2019fault}
G.~Iannace, G.~Ciaburro, and A.~Trematerra, ``Fault diagnosis for uav blades
  using artificial neural network,'' \emph{Robotics}, vol.~8, no.~3, p.~59,
  2019.

\bibitem{zhang2018deep}
W.~Zhang, C.~Li, G.~Peng, Y.~Chen, and Z.~Zhang, ``A deep convolutional neural
  network with new training methods for bearing fault diagnosis under noisy
  environment and different working load,'' \emph{Mechanical Systems and Signal
  Processing}, vol. 100, pp. 439--453, 2018.

\bibitem{chen2017deep}
Z.~Chen, S.~Deng, X.~Chen, C.~Li, R.-V. Sanchez, and H.~Qin, ``Deep neural
  networks-based rolling bearing fault diagnosis,'' \emph{Microelectronics
  Reliability}, vol.~75, pp. 327--333, 2017.

\bibitem{de2016railway}
T.~de~Bruin, K.~Verbert, and R.~Babu{\v{s}}ka, ``Railway track circuit fault
  diagnosis using recurrent neural networks,'' \emph{IEEE transactions on
  neural networks and learning systems}, vol.~28, no.~3, pp. 523--533, 2016.

\bibitem{pereira2016brain}
S.~Pereira, A.~Pinto, V.~Alves, and C.~A. Silva, ``Brain tumor segmentation
  using convolutional neural networks in mri images,'' \emph{IEEE transactions
  on medical imaging}, vol.~35, no.~5, pp. 1240--1251, 2016.

\bibitem{levi2015age}
G.~Levi and T.~Hassner, ``Age and gender classification using convolutional
  neural networks,'' in \emph{Proceedings of the iEEE conference on computer
  vision and pattern recognition workshops}, 2015, pp. 34--42.

\bibitem{ma2016learning}
L.~Ma, Z.~Lu, and H.~Li, ``Learning to answer questions from image using
  convolutional neural network,'' in \emph{Thirtieth AAAI Conference on
  Artificial Intelligence}, 2016.

\bibitem{wen2017new}
L.~Wen, X.~Li, L.~Gao, and Y.~Zhang, ``A new convolutional neural network-based
  data-driven fault diagnosis method,'' \emph{IEEE Transactions on Industrial
  Electronics}, vol.~65, no.~7, pp. 5990--5998, 2017.

\bibitem{guo2018hybrid}
D.~Guo, M.~Zhong, H.~Ji, Y.~Liu, and R.~Yang, ``A hybrid feature model and deep
  learning based fault diagnosis for unmanned aerial vehicle sensors,''
  \emph{Neurocomputing}, vol. 319, pp. 155--163, 2018.

\bibitem{wen}
L.~Wen, L.~Gao, and X.~Li, ``A new deep transfer learning based on sparse
  auto-encoder for fault diagnosis,'' \emph{IEEE Transactions on Systems, Man,
  and Cybernetics: Systems}, vol.~49, no.~1, pp. 136--144, 2017.

\bibitem{zhang2017transfer}
R.~Zhang, H.~Tao, L.~Wu, and Y.~Guan, ``Transfer learning with neural networks
  for bearing fault diagnosis in changing working conditions,'' \emph{IEEE
  Access}, vol.~5, pp. 14\,347--14\,357, 2017.

\bibitem{cao2018preprocessing}
P.~Cao, S.~Zhang, and J.~Tang, ``Preprocessing-free gear fault diagnosis using
  small datasets with deep convolutional neural network-based transfer
  learning,'' \emph{IEEE Access}, vol.~6, pp. 26\,241--26\,253, 2018.

\bibitem{glowacz2018early}
A.~Glowacz, W.~Glowacz, Z.~Glowacz, and J.~Kozik, ``Early fault diagnosis of
  bearing and stator faults of the single-phase induction motor using acoustic
  signals,'' \emph{Measurement}, vol. 113, pp. 1--9, 2018.

\bibitem{kemalkar2016engine}
A.~K. Kemalkar and V.~K. Bairagi, ``Engine fault diagnosis using sound
  analysis,'' in \emph{2016 International Conference on Automatic Control and
  Dynamic Optimization Techniques (ICACDOT)}.\hskip 1em plus 0.5em minus
  0.4em\relax IEEE, 2016, pp. 943--946.

\bibitem{sajedi2019intensity}
{S.O. Sajedi and X. Liang}, ``Intensity-based feature selection for near
  real-time damage diagnosis of building structures,'' \emph{arXiv preprint
  arXiv:1910.11240}, 2019.

\bibitem{sajedi2019vibration}
S.~Sajedi and X.~Liang, ``Vibration-based semantic damage segmentation for
  large-scale structural health monitoring,'' \emph{Computer-Aided Civil and
  Infrastructure Engineering}, 2019.

\bibitem{liang2018simulation}
X.~Liang, K.~M. Mosalam, and S.~Muin, ``Simulation-based data-driven damage
  detection for highway bridge systems,'' in \emph{11th National Conference on
  Earthquake Engineering}, 2018.

\bibitem{sajedi2020data}
{S.O. Sajedi and X. Liang}, ``A data-driven framework for near real-time and
  robust damage diagnosis of building structures,'' \emph{Structural Control
  and Health Monitoring}, vol.~27, no.~3, p. e2488, 2020.

\bibitem{pan2009survey}
S.~J. Pan and Q.~Yang, ``A survey on transfer learning,'' \emph{IEEE
  Transactions on knowledge and data engineering}, vol.~22, no.~10, pp.
  1345--1359, 2009.

\end{thebibliography}

\end{document}